\documentclass[]{aa} 
\usepackage{graphicx}
\usepackage{txfonts}

\begin{document}
   \title{Spitzer observations of spacecraft target 162173 (1999 JU3)}
   \author{H. Campins
          \inst{1,2}
	  \and
	     J. P. Emery
	  \inst{3}
	  \and
	  M. Kelley
	  \inst{4}
	  \and
	  Y. Fern\'andez
	  \inst{1}
	  \and
	  J. Licandro
	  \inst{2}
	  \and
	  M. Delb\'o
	  \inst{5}
	  \and
	  A. Barucci
	  \inst{6}
	  \and
	  E. Dotto
	  \inst{7}}

   \offprints{H. Campins}

  \institute{Instituto de Astrof\'{\i}sica de Canarias, c/V\'{\i}a L\'actea s/n, 38200 La Laguna, Tenerife, Spain. \\
              \email{campins@physics.ucf.edu}
              \and
              Physics Department, University of Central Florida, Orlando, FL, 32816, USA.\\
	        \and
	        University of Tennessee, Knoxville, Tennessee.\\
	        \and
	        Department of Astronomy, University of Maryland, College Park, MD 20742-2421, USA.\\
              \and
              UNS, CNRS, Observatoire de la C\^ote d'Azur, Nice, France.\\
              \and
              LESIA, Observatoire de Paris, France.\\
              \and
              INAF, Osservatorio Astronomico di Roma, Roma, Italy. \\
 }
   \date{Accepted for publication as a Letter in Astronomy and Astrophysics}


  \abstract
   {Near-Earth asteroid 162173 (1999 JU3) is the primary target of the Japanese Aerospace Exploration
   Agency (JAXA) Hayabusa-2 sample return mission, and is also on the list of potential targets for the
   European Space Agency (ESA) Marco Polo sample return mission.  Earth-based studies of this object are
   fundamental to these missions.}
   {Our aim is to provide new constraints on the surface properties of this asteroid.  }
   {We present a mid-infrared spectrum (5-38 $\mu$m) obtained with NASA's Spitzer Space Telescope in May 2008 and results from the application of thermal models.}
   {These observations place new constraints on the surface properties of this asteroid. To fit our spectrum we used the near-Earth asteroid thermal model (NEATM) and the more complex thermophysical model (TPM). However, the position of the spin-pole, which is uncertain, is a crucial input parameter for constraining the thermal inertia with the TPM; hence, we consider two pole orientations. First is the extreme case of an equatorial retrograde geometry from which we derive a rigorous lower limit to the thermal inertia of 150 $Jm^{-2} s^{-0.5} K^{-1}$. Second, when we adopt the pole orientation of Abe et al. (2008a) our best-fit thermal model yields a value for the thermal inertia of 700 $\pm$ 200 $Jm^{-2} s^{-0.5} K^{-1}$ and even higher values are allowed by the uncertainty in the spectral shape due to the absolute flux calibration. Our best estimates of the diameter (0.90 $\pm$ 0.14 km) and geometric albedo (0.07 $\pm$ 0.01) of asteroid 162173 are consistent with values based on previous mid-infrared observations.
   }
   {We establish a rigorous lower limit to the thermal inertia, which is unlikely but possible, and would be consistent with a fine regolith similar to wthat is found for asteroid 433 Eros. However, the thermal inertia is expected to be higher, possibly similar to or greater than that on asteroid 25143 Itokawa.  An Accurately determining the spin-pole of asteroid 162173 will narrow the range of possible values for its thermal inertia.}

   \keywords{asteroids - infrared - spectroscopy}
   \titlerunning{Spitzer Observations of 162173 (1999 JU3)}
   \maketitle


\section{Introduction}
Near-Earth asteroid 162173 (1999 JU3) is the primary target of the Japanese Aerospace Exploration Agency (JAXA) Hayabusa-2 sample return mission.  This object is also on the list of potential targets for the European Space Agency (ESA) Marco Polo mission, which was selected for an assessment phase study in late 2007. The goal of both missions is to return to Earth samples of a primitive near-Earth asteroid for laboratory analysis.

Since these space missions seek to understand the origin and nature of volatile and organic material in the early Solar System, samples of a primitive asteroid (C, P, or D asteroids in the Tholen classification system; Tholen and Barucci \cite{Tholen1989}) are highly desirable.  Asteroid 162173 is a C-class object, more specifically a Cg-type in the SMASS classification system (Binzel et al. \cite{Binzeletal01}). In the asteroid population, C-class objects are abundant in the middle to outer Main Belt, and are primitive, volatile-rich remnants of the early solar system. Spectroscopy of C-class objects and comparisons with potential meteorite analogs suggest surface constituents of anhydrous silicates, hydrated minerals, organic polymers, magnetite, and sulfides (e.g., Rivkin et al. \cite{Rivkingetal02}; Gaffey et. al. \cite{Gaffeyetal02}).  Dynamically, asteroid 162173 is most likely to have come from the main asteroid belt through the 3:1 jovian mean motion resonance (e.g., Bottke et al. \cite{Bottkeetal02}).  At visible and near infrared wavelengths (0.5 to 2.0 $\mu$m), this asteroid shows a spectrum with a neutral to bluish spectral slope (Binzel et al. \cite{Binzeletal01}; Abe et al., \cite{Abeatal08}).  A broad and shallow absorption, centered near 0.7 $\mu$m and associated with hydrated minerals (iron-bearing phyllosilicates), has been reported at one rotational phase by Vilas (\cite{Vilas08}).  This asteroid appears to be nearly spherical with a rotational amplitude of about 0.1 magnitudes (Abe et al. \cite{Abeatal08}) and a rotation period of 0.3178 days (Hasegawa et al. \cite{Hasegawaetal08}).  Ground-based and space-based narrowband mid-IR photometry were used by Hasegawa et al. (\cite{Hasegawaetal08}) to derive estimates of the geometric albedo (0.063 +0.020 -0.015), diameter (0.91 $\pm$ 0.12 km), and thermal inertia ($\gtrsim $ 500 $Jm^{-2} s^{-0.5} K^{-1}$). The small rotational amplitude makes determining a spin-pole orientation for this asteroid a challenging task; nevertheless, a spin-pole orientation has been estimated at ecliptic longitude and latitude of 331 and 20$^{\circ} $, respectively (Abe et al. \cite{Abeatal08a}).  In this work, we present new mid-infrared spectroscopy (5-38 $\mu$m) obtained with NASA's Spitzer Space Telescope, which has allowed us to place new constraints on the surface properties of this asteroid.

\section{Observations and data reduction}

The spectrum was obtained with the Infrared Spectrograph (IRS) instrument (Houck et al. \cite{Houcketal04}) on NASA's {\em Spitzer} Space Telescope (Werner et al. \cite{Werneretal04}), on UT May 2.084, 2008.  The asteroid was observed at a heliocentric distance of 1.202 AU, at a Spitzer distance of 0.416 AU, a phase angle of 52.6$^{\circ} $, and a galactic latitude of 3.2$^{\circ} $.  The IRS measures spectra over the range 5.2-38 $\mu$m. The observations presented here used the low spectral resolution mode ($R = \lambda / \Delta \lambda \sim$ 64-128), which covers the full spectral range in four long-slit segments, and the final spectrum corresponds to a total on-object exposure time of 738.2 seconds. These four segments are short wavelength, 2nd order (SL2; 5.2-8.5 $\mu$m), short wavelength, 1st order (SL1; 7.4-14.2 $\mu$m), long wavelength, 2nd order (LL2; 14.0-21.5 $\mu$m), and long wavelength, 1st order (LL1; 19.5-38.0 $\mu$m). When each of the low-resolution orders is centered on the target, the other is offset ($\sim$ 50" for SL and $\sim$108" for LL) from the target, providing a measurement of the sky background.  We anticipated a crowded field, so we also obtained the sky background from a ``shadow'' observation, which duplicated the first observation but with the asteroid out of the field of view. We did not use the peak-up routine because (a) the ephemeris uncertainty was only about 0.3 arcsec, much less than the slit widths, (b) the Spitzer pointing accuracy is about 1 arcsec, also less than the slit widths, and (c) the object's Galactic latitude was low so we did not want IRS to inadvertently peak-up on a background source. Our spectra were cleaned, extracted, and co-added using the Spitzer IRS Custom Extraction tool (SPICE), and our own routines. More details on our data reduction can be found in Kelley et al. (\cite{Kelleyetal09}).

When the reduction was completed, we found a discrepancy of 10\% between the fluxes of overlapping wavelengths in the SL and LL orders, which is equal to the expected uncertainty in the absolute flux calibration (Houck et al. \cite{Houcketal04}), and small offsets like this one are not unusual\footnote {We considered and ruled out other potential sources for the SL and LL flux difference, such as placement of the object within the slit and rotational variability of the asteroid.}.  We scaled the LL and SL fluxes to match each other in the overlapping wavelengths, the SL flux by 1.05 and the LL by 0.95, and we include this 10\% systematic absolute calibration uncertainty in the modeled parameters discussed in the next section. This spectrum is plotted in Fig. 1 along with the best-fit thermophysical model (TPM, Sect. 3).



\begin{figure}
	\centering
	\includegraphics[width=\columnwidth]{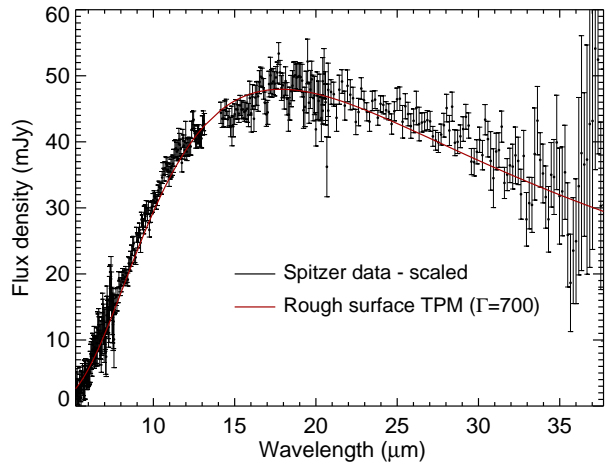}
	\caption{The Spitzer spectrum of near-Earth asteroid 162173 (1999 JU3) plotted along with our best-fit thermophysical model (TPM).  The model parameters are given in Table 1 and discussed in Sect. 3. The LL and SL fluxes have been scaled to match each other in the overlapping wavelengths (see Sect. 2).}
 	\label{Fig1}
 \end{figure}

\section{Thermal modeling}

The flux from near-Earth asteroids in the wavelength range covered by the IRS is dominated by thermal emission. The measured spectral energy distribution (SED) depends on the object's size, composition, and temperature distribution. This last term is dependent on several factors, including distance from the Sun, albedo, thermal inertia, surface roughness, rotation rate, shape, and spin-pole orientation. We used two models to fit our 5 to 38 $\mu$m spectrum, the relatively simple near-Earth asteroid thermal model (NEATM, Harris \cite{Harris1998}) and the more complex thermophysical model (TPM, e.g., Harris and Lagerros \cite{Harris02} and references therein).

The NEATM is a refinement of the standard thermal model (STM, Lebofsky et al. \cite{Lebofskyetal1986}; Lebofsky and Spencer \cite{Lebofskyetal1989}), which was developed and calibrated for main-belt asteroids.  The NEATM accounts for observations at larger phase angles and contemplates possible deviations from the STM in the thermal behavior of smaller asteroids.  Unlike the STM, the NEATM requires observations at multiple wavelengths and uses this information to force the model temperature distribution to be consistent with the apparent color temperature of the asteroid. The NEATM solves simultaneously for the beaming parameter ($\eta$) and the diameter ($D$). The beaming parameter was originally introduced in the STM to allow the model temperature distribution to fit the observed enhancement of thermal emission at small solar phase angles due to surface roughness. In practice, $\eta$ can be thought of as a modeling parameter that allows a first-order correction for any effect that influences the observed surface temperature distribution (such as beaming, thermal inertia, and rotation). The values of $\eta$ derived for asteroid 162173 using the NEATM fits to our spectrum are given in Table 1; for all our thermal models we assumed a bolometric emissivity ($\epsilon$) of 0.9. These $\eta$ values can be compared with those of other asteroids studied in a similar manner, and in fact fit well the trend of increasing $\eta$ (lower thermal fluxes and cooler color temperatures) with increasing solar phase angle, observed for about 40 near-Earth asteroids so far (e.g., Delb\'o \cite{Delbo04}, Campins et al. \cite{Campinsetal09}). We did not make any observations of this asteroid's reflected continuum, so to estimate the geometric albedo ($p_V$), we adopted the values given by Hasegawa et al. (\cite{Hasegawaetal08}) of $H_V$ = 18.81$\pm$0.03 and $G$ = -0.115$\pm$0.009, where $H_V$ is the visible absolute magnitude and $G$ is the slope parameter.

\begin{table*}
\centering
\caption{NEATM and TPM best-fit Models}
\begin{tabular}{l l c c c c}
\hline\hline

Model &Scaling & Diameter & $\eta$ & $\Gamma$  & $p_V$ \\
& & (km) & & ($Jm^{-2} s^{-0.5} K^{-1}$) & \\ \hline \hline
NEATM & no scaling & 0.97 $\pm$ 0.15 & 1.90 $\pm$ 0.17 & & 0.06 $\pm$ 0.01 \\  \hline
NEATM & scaled orders & 0.91 $\pm$ 0.14 & 1.63 $\pm$ 0.15  & & 0.07 $\pm$ 0.01 \\  \hline
TPM1 & scaled orders & 0.87 $\pm$ 0.13 & & 150 & 0.07 $\pm$ 0.01 \\  \hline
TPM2  & no scaling & 0.97 $\pm$ 0.15 & & $\sim$1500 (Fig. 2) & 0.06 $\pm$ 0.01 \\  \hline
TPM2 & scaled orders & 0.90 $\pm$ 0.14 & & 700 $\pm$ 200 & 0.07 $\pm$ 0.01 \\  \hline
\end{tabular}

\label{Table1}
\end{table*}

 In Table 1, scaling means that a correction was applied to the fluxes in the LL and SL orders (Sect. 2).  $Diameter$, $\eta$, $\Gamma$, and $p_V$ are the derived diameter, beaming parameter, thermal inertia, and geometric albedo, respectively. TPM1 yields a rigorous lower limit to the thermal inertia (Sect. 3) and TPM2 assumes the pole orientation of Abe et al. (\cite{Abeatal08a}) given in Sect. 1. Uncertainties in the diameter estimated with thermal modeling usually exceed the formal errors and are typically 10-15\% (e.g., Delb\'o et al. \cite{Delboetal03}). The listed error in the geometric albedo includes the uncertainty in $H_V$, with a 0.05 magnitude component due to the rotational amplitude of 0.1 magnitudes, and the diameter uncertainty propagated into the albedo.

The TPM is applied to asteroids with known shape and spin state, and is generally used to constrain the thermal inertia ($\Gamma$), which is particularly relevant to the proposed sample-return missions since $\Gamma$ can be diagnostic of the regolith. For example, typical values of  $\Gamma$ (in S.I. units $Jm^{-2} s^{-0.5} K^{-1}$) are 50 for lunar regolith, 400 for coarse sands, and 2500 for bare solid rocks (e.g., Delb\'o and Tanga \cite{Delbo09}, and references therein). As mentioned in Sect. 1, there are estimates of the rotation period, approximate spin-pole orientation and the shape of asteroid 162173, so in principle we can apply the TPM.  However, because of the uncertainties in the pole position, the TPM does not uniquely constrain the thermal inertia of this asteroid; we elaborate on this point in the next paragraph.  Another input parameter needed in this case is the surface roughness of the asteroid. This roughness can also be constrained by the TPM if observations are made at several phase angles, but in our case it has to be assumed because we obtained only one spectrum. Little is known about the typical surface roughness of NEAs, and there is no consensus about what roughness parameters should be used (e.g., Harris et al \cite{Harrisetal07} and references therein). We explored the effect of surface roughness by incorporating two values of surface roughness into our models (see below).  The difference in the derived thermal inertia between the smooth surface case and the rough case was less than 30\%.  Because this uncertainty is considerably smaller than that due to the spin-pole orientation, it is not meaningful to try to refine the roughness parameters at this time.

As mentioned, the spin-pole orientation is a crucial input parameter for constraining the thermal inertia. However, the estimate of the spin-pole position for asteroid 162173 is uncertain because the amplitude of the rotational light curve is small and the range of geometries for the light curve observations has been limited. We, therefore, consider two cases for the orientation of the spin-pole. The first is the extreme case of an equatorial retrograde geometry, i.e., the spin-pole pointing at the South ecliptic pole.  As input parameters for this TPM, we used a spherical shape (a reasonable assumption given the small amplitude of the light curve), the equatorial retrograde geometry, the rotation period of 0.3178 days, and a perfectly smooth surface (this last parameter yields a lower thermal inertia than in the rough surface case).  Then we allowed the unconstrained parameters in the model ($D$, $p_V$, and  $\Gamma$) to vary, an approach that worked particularly well because our spectrum spans a broad wavelength range that includes the peak in thermal emission.  We generated a grid of models with  $\Gamma$ ranging from 0 to 2500 $Jm^{-2} s^{-0.5} K^{-1}$, and the best-fit model was determined by minimizing the reduced $\chi ^2$ between the model fluxes and our data. This case (equatorial retrograde geometry and smooth surface) yields a rigorous lower limit to the thermal inertia of 150 $Jm^{-2} s^{-0.5} K^{-1}$.

For the second case, we adopted the pole orientation of Abe et al. (\cite{Abeatal08a}) given in Sect. 1, which implies a subsolar latitude of approximately 30 $^{\circ} $ at the time of the Spitzer observations. As in the first case, we generated a grid of models, and the best-fit model was determined by minimizing the reduced $\chi ^2$.  In Fig. 1, we plot our best-fit TPM model along with the observed spectrum, and the corresponding parameters are listed in Table 1, again assuming $\epsilon$ = 0.9 and the $H_V$ and $G$ from Hasegawa et al. (\cite{Hasegawaetal08}). Fig. 2 shows the reduced  $\chi ^2$ vs thermal inertia and the estimated diameter and geometric albedo vs thermal inertia for the rough-surface case (the parameters for the rough-surface model are $\gamma$ = 53$^{\circ} $ [half-angle opening of spherical section craters] and f = 0.77 [fractional surface coverage of craters]). The smooth surface model is not plotted because of space limitations, but fits equally well and yields about 30\% lower values for the thermal inertia.

Uncertainties in thermal modeling usually exceed the formal errors resulting from the scatter of the flux measurements. The main sources of uncertainties in our derived physical parameters are the absolute photometric uncertainty in the thermal flux (diameter), the relative spectral uncertainty (temperature, beaming parameter and thermal conductivity), and the uncertainty in $H_V$ (albedo), including the rotational light curve.  Comparison of diameters derived from thermal models with those derived from other sources, such as radar, indicates that overall errors are normally less than 15\% in diameter (Delb\'o et al. \cite{Delboetal03}). The contribution of relative spectral uncertainty to errors in the thermal continuum and derived parameters is dominated by the absolute-flux-calibration adjustment of the LL and SL segments of the spectrum (Sect. 2). The estimated uncertainties are given in Table 1, and in the case of the (second case) TPM, they are also illustrated in Fig. 2.

Spectral features superposed on the mid-infrared thermal continuum have been detected in other asteroids and can be compositionally diagnostic. However, except for a possible weak feature near 15 $\mu$m (Fig. 1), no obvious structure is apparent in our spectrum, and a detailed analysis is deferred to a later publication.

\begin{figure}
	\centering
	\includegraphics[width=\columnwidth]{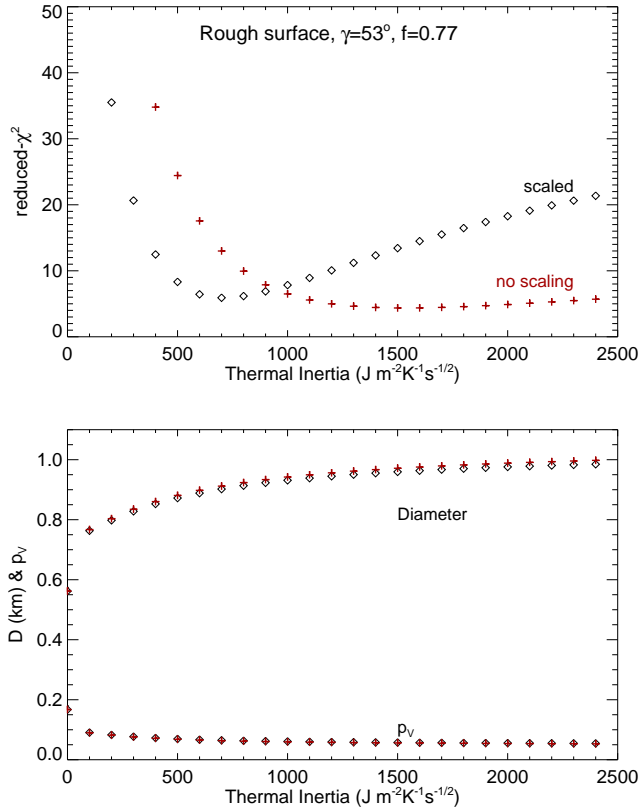}
	\caption{The the reduced  $\chi ^2$ vs thermal inertia for a rough surface TPM.  The parameters for the rough surface model are given in Sect. 3.  The bottom plot shows size and geometric albedo vs thermal inertia. The best-fit model was determined by minimizing the reduced  $\chi ^2$ between the model and the data, and is shown in Figure 1 along with the observed spectrum.}
 	\label{Fig2}
 \end{figure}

\section{Results and conclusion}

Our results are summarized in Table 1, and the most relevant to the proposed sample return missions are the constraints our observations impose on the thermal inertia, $\Gamma$. When we adopt the extreme case of an equatorial retrograde geometry and a smooth surface for this asteroid, we derive a rigorous lower limit to the thermal inertia of 150 $Jm^{-2} s^{-0.5} K^{-1}$. This value would be consistent with a fine regolith, similar to the one found in asteroid 433 Eros.  Adopting the pole orientation of Abe et al. (\cite{Abeatal08a}) yields a best-fit model with thermal inertia at 700 +/- 200  $Jm^{-2} s^{-0.5} K^{-1}$. Even higher values are allowed by the uncertainty in the spectral shape due to absolute flux calibration. This best-fit thermal inertia and its size place asteroid 162173 about two-sigma above the trend toward increasing thermal inertia with decreasing asteroid diameter as discussed by Delb\'o et al. (\cite{Delboetal07}).  The range in possible thermal inertias for asteroid 25143 Itokawa adopted by Delb\'o et al. (\cite{Delboetal07}) (based on M{\"u}ller et al. \cite{Mulleretal05} and Mueller \cite{Mueller07}) is 350 to 800 $Jm^{-2} s^{-0.5} K^{-1}$, which is within the range of possible values for asteroid 162173.  If the spin-pole of Abe et al. (\cite{Abeatal08a}) turns out to be correct, these last two results would suggest that the surface of asteroid 16217 has a texture rockier than or similar to that of asteroid 25143 Itokawa.  However, given the uncertainties in the spin-pole position, we hesitate to even make this analogy.

Our estimates of the diameter and albedo of asteroid 162173 are consistent with those of Hasegawa et al. (\cite{Hasegawaetal08}). However, there are marginally significant discrepancies that may deserve further attention. There are temperature differences between their space-based (Akari) and ground-based (Subaru) data sets and between their data and ours.  As a way to illustrate these differences we give our best-fit NEATM parameters for their Akari and Subaru data, which are $D$ = 0.80$\pm$0.12 km, $p_V$ = 0.08$\pm$0.03, $\eta$=1.0$\pm$0.4 and $D$ =1.13$\pm$0.17 km, $p_V$ = 0.04$\pm$0.02, $\eta$ = 2.1$\pm$0.6, respectively. Only for $\eta$ are the differences slightly larger than one sigma.  It may be possible to constrain the spin-pole orientation better by modeling all three data sets (ours and the two sets in Hasegawa et al. \cite{Hasegawaetal08}) simultaneously, allowing the pole solution to be a free parameter; however, this is work beyond the scope of this letter.

In summary, thermal models of our Spitzer spectrum of asteroid 162173 (1999 JU3) provide a rigorous lower limit to the thermal inertia of 150 $Jm^{-2} s^{-0.5} K^{-1}$, which is unlikely but possible. This value would be consistent with a fine regolith, similar to the one found in asteroid 433 Eros. However, if we adopt the pole orientation of Abe et al. (\cite{Abeatal08a}), the best-fit model yields a thermal inertia of 700 $\pm$ 200  $Jm^{-2} s^{-0.5} K^{-1}$. Even higher values are allowed by the uncertainty in the spectral shape due to absolute flux calibration. Our best estimates of the diameter (0.90 $\pm$ 0.14 km) and geometric albedo (0.07 $\pm$ 0.02) of asteroid 162173 are consistent with those of Hasegawa et al. (\cite{Hasegawaetal08}).

\begin{acknowledgements}
HC acknowledges support from NASA's Spitzer Science Center, Jet Propulsion Laboratory, and Planetary Astronomy program.  HC was a visiting Fulbright Scholar at the ``Instituto de Astrof\'{\i}sica de Canarias'' in Tenerife, Spain. JL acknowledges support from the Spanish ``Ministerio de Ciencia e Innovaci\'on'' projects AYA2005-07808-C03-02 and AYA2008-06202-C03-02.  We also thank an anonymous referee for helpful comments.
\end{acknowledgements}

\end{document}